\documentclass[conference]{IEEEtran}

\IEEEoverridecommandlockouts
\usepackage{cite}
\usepackage{amsmath,amssymb,amsfonts}
\usepackage{algorithmic}
\usepackage{graphicx}
\usepackage{acronym}
\usepackage{textcomp}
 \usepackage{balance}
\usepackage{wrapfig}
 \usepackage{booktabs}

\usepackage[font=footnotesize]{caption}

\usepackage{subcaption}

\usepackage{xcolor}
\usepackage{tabularray}
\usepackage{multirow}

\usepackage{eurosym}
\usepackage{tabularx}
\def\BibTeX{{\rm B\kern-.05em{\sc i\kern-.025em b}\kern-.08em
    T\kern-.1667em\lower.7ex\hbox{E}\kern-.125emX}}
\usepackage[hyphens]{url}
\usepackage[hidelinks]{hyperref}
\hypersetup{breaklinks=true}
\begin{document}
\title{From Individual Consumers to Energy Communities: A Techno-economic Assessment of Swiss Local Electricity Communities
\thanks{This research has received funding from the European Union’s Horizon Europe research and innovation programme under Grant Agreement for Project Nº 101235482 and the Swiss Secretariat for Education, Research, and Innovation (SERI) under contract Nº25.00325.}}

\author{
\IEEEauthorblockN{Na Li, Binod Koirala}
\IEEEauthorblockA{Urban Energy Systems Laboratory\\
Empa, Swiss Federal Laboratories for Materials Science and Technology\\
Überlandstrasse 129, 8600 Dübendorf, Switzerland}
\{na.li, binod.koirala\}@empa.ch
}

\maketitle

\begin{abstract}
As energy communities move from policy design to implementation in Switzerland, understanding their performance in practice has become increasingly important. A techno-economic assessment of a regulation-compliant~\ac{LEC} is presented under the new Swiss legal framework in this study. A reference case without local electricity exchange is compared to a~\ac{LEC} scenario with internal electricity sharing. Results show that~\ac{LEC} participation increases local renewable utilization, reduces grid exports, and delivers economic benefits to both consumers and prosumers. A sensitivity analysis further indicates that internal electricity pricing plays a critical role in shaping trade-offs between overall efficiency and fairness in benefit distribution. This exploratory study provides practical insights to support informed decision-making and the future development of~\ac{LEC} in Switzerland.

\end{abstract}
\acrodef{RES}{Renewable Energy Sources}
\acrodef{LEC}{Local Electricity Community}
\acrodef{DSO}{Distribution System Operator}
\acrodef{PV}{Photovoltaic}
\acrodef{CV}{Coefficient of Variation}


\section{Introduction}

\IEEEPARstart{E}{nergy} communities have increasingly been recognized as an institutional mechanism for integrating distributed~\acp{RES} into the electric power systems while enabling active participation of local actors~\cite{li2023economic, minuto2022energy}. Across Europe, collective self-consumption schemes~\cite{giarmana2023managing} have been incorporated into national regulatory frameworks following the adoption of the Renewable Energy Directive (RED II)~\cite{european2018red} and the Clean Energy for All Europeans Package~\cite{european2019clean}, which formally recognize energy communities as legitimate actors in the electricity sector. Together, these regulatory developments reflect a structural shift from centrally coordinated generation towards more decentralized and participatory energy systems, in which households and small-scale producers are no longer passive consumers but increasingly engage in electricity production, sharing, and local coordination.

Within the European landscape, Switzerland has followed a gradual and path-dependent approach to enabling collective self-consumption. Early regulatory instruments such as the Zusammenschluss zum Eigenverbrauch (ZEV) (2018) and its virtual extension (vZEV) (2025) allowed electricity sharing among participants located behind a single grid connection point or connected to the same connection point, as regulated under the Energy Act (EnG)~\cite{swiss2016EnG}. Building on these arrangements, the revised Electricity Supply Act (Stromversorgungsgesetz, StromVG) introduces the Local Electricity Community (\ac{LEC}), entering into force on 1 January 2026~\cite{swiss2007stromVG}. In contrast to ZEV and vZEV, the~\ac{LEC} framework allows electricity exchange among participants connected to different grid connection points within the service area of the same~\ac{DSO} and within the same municipality, while granting predefined reductions in network usage charges for locally exchanged electricity~\cite{swiss2008stromVV, swiss2025LEG}.

A growing body of literature has examined energy communities from technical, economic, and socio-institutional perspectives, including business models and value propositions~\cite{koirala2016energetic, reis2021business}, benefit-sharing rules under different ownerships and demand-profile conditions~\cite{minuto2022energy}, cost allocation mechanisms~\cite{li2021cost, li2023economic}, and peer-to-peer trading schemes~\cite{hahnel2020becoming}.
Beyond techno-economic analyses, some literature investigates social acceptance~\cite{li2021socail}, governance structures~\cite{lowitzsch2020renewable}, and participation and organizational arrangements of energy communities~\cite{koirala2018trust, gjorgievski2021social}. 

These analyses, however, primarily concentrate on the design and comparison of alternative business models, sharing mechanisms, and allocation rules for energy communities. By contrast, the Swiss~\ac{LEC} fixes the institutional framework for local electricity exchange through regulation and shifts the analytical focus towards operational and internal decision-making within a predefined rule set. Quantitative analysis of system-level performance, household-level outcomes, and interactions with the distribution grid under these conditions remains limited.

To address this gap, this paper presents a quantitative case study based on real-world consumption and~\ac{PV} generation data from the NEST platform and vertical energy district, and analyzes a~\ac{LEC}-compliant local electricity exchange scenario in accordance with the new Swiss regulatory framework. We evaluate community-level impacts on self-consumption and grid interaction, as well as household-level economic outcomes, and examine how internal electricity exchange prices shape benefit distribution under the~\ac{LEC} framework. The main contributions of this work are threefold: (i) the development of a transparent and rule-consistent settlement and accounting model following the Swiss~\ac{LEC} regulation; (ii) a quantitative assessment of both community-level and household-level impacts using real-world electricity consumption and~\ac{PV} generation data; and (iii) an explicit analysis of how internal electricity pricing influences the distribution of economic benefits among households.

The remainder of this paper is organized as follows: Section~\ref{modeling} introduces the~\ac{LEC} regulatory framework, settlement model, allocation rules, and performance indicators. Section~\ref{Results analysis} presents results, followed by a discussion of practical implications. Finally, Section~\ref{Conclusions} concludes the paper by providing future work recommendations.

\section{Regulatory-Constrained Modeling Framework and Case Study Design} \label{modeling}
\subsection{Swiss regulatory framework and LEC settlement structure}

\acp{LEC} are regulated under the Swiss Electricity Supply Act (StromVG) and the corresponding ordinance (StromVV), which define the spatial, technical, and accounting conditions for local electricity exchange. A~\ac{LEC} may consist of consumers, prosumers with~\ac{PV} generation, and storage operators (optional). All participants must be located within the same municipality, connected to the same~\ac{DSO} and at the same network level.

Participation in a~\ac{LEC} is voluntary, and each~\ac{LEC} is required to appoint a representative responsible for coordinating the community and acting as the single point of contact with the~\ac{DSO}. All households remain individually metered by the~\ac{DSO}, and locally exchanged electricity is identified based on 15-minute metering data provided by the~\ac{DSO}. Local electricity exchange is physically delivered via the public distribution network and settled separately from residual grid imports and exports.

In line with the regulatory framework, electricity exchanged within a~\ac{LEC} benefits from a reduced network usage tariff. A reduction of 40\% applies when local exchange occurs without a change in voltage level, and 20\% otherwise.

\subsection{Internal pricing and allocation rules} \label{Internal pricing and allocation rules}
Under the Swiss regulatory framework (StromVG), local electricity exchange within a~\ac{LEC} is settled at an internal price $\lambda^{\mathrm{loc}}$, which is agreed upon by community members, without a prescribed pricing rule. In this work, we assume that the local exchange price lies between the feed-in remuneration for exported electricity, $\lambda^{\mathrm{export}}$, and the retail energy price $\lambda^{\mathrm{E}}$. This assumption reflects a fairness principle and ensures that both consuming and supplying households benefit economically from local electricity exchange.

Locally exchanged electricity is allocated to households with residual demand in proportion to their electricity consumption. Correspondingly, revenues from local electricity exchange are attributed to~\ac{PV}-owning households in proportion to their surplus~\ac{PV} generation. This proportional allocation ensures energy conservation at the community level and provides a transparent and neutral settlement rule, avoiding preferential treatment or strategic behavior.

\subsection{LEC system modeling and case definitions}\label{LEC model}
This subsection formalizes the~\ac{LEC} system model based on the settlement rules and electricity prices defined in Section~\ref{Internal pricing and allocation rules}. The model is formulated at both the household and community level and distinguishes between a reference case without local electricity exchange and a~\ac{LEC} case with local electricity exchange.

Let $P^{\mathrm{con}}_{i,t}$ denote the electricity consumption of household $i$ during time interval $t$, and $P^{\mathrm{gen}}_{i,t}$ its~\ac{PV} generation if installed. In both cases,~\ac{PV} generation is first self-consumed at the household level:
\begin{equation}
P^{\mathrm{self}}_{i,t}
= \min\!\left(P^{\mathrm{con}}_{i,t},\, P^{\mathrm{gen}}_{i,t}\right).
\end{equation}

\subsubsection*{Reference Case (no~\ac{LEC})}

In the reference case, households operate individually without local electricity exchange. Residual electricity demand is supplied from the grid, and surplus PV generation is exported to the grid:
\begin{align}
P^{\mathrm{ref,import}}_{i, t}
&= P^{\mathrm{con}}_{i, t} - P^{\mathrm{self}}_{i, t},\\
P^{\mathrm{ref,export}}_{i, t}
&= P^{\mathrm{gen}}_{i, t} - P^{\mathrm{self}}_{i, t}.
\end{align}

The energy cost of household $i$ over the analysis horizon $T$ in the reference case is given by
\begin{equation}
C^{\mathrm{ref}}_{i}
= \sum_{t=1}^{T}
\left(
P^{\mathrm{ref,import}}_{i, t}\,\lambda^{\mathrm{import}}
-
P^{\mathrm{ref,export}}_{i, t}\,\lambda^{\mathrm{export}}
\right),
\end{equation}
where $\lambda^{\mathrm{import}}$ and $\lambda^{\mathrm{export}}$ denote the retail electricity price and feed-in remuneration.

\subsubsection*{~\ac{LEC} case}

Under the~\ac{LEC} framework, only surplus~\ac{PV} generation after self-consumption is made available for local electricity exchange:
\begin{equation}
P^{\mathrm{sur}}_{i, t}
= P^{\mathrm{gen}}_{i, t} - P^{\mathrm{self}}_{i, t}.
\end{equation}

Community-level surplus generation and residual demand are defined as
\begin{align}
P^{\mathrm{LEC,sur}}_{t}
&= \sum_i P^{\mathrm{sur}}_{i, t},\\
P^{\mathrm{LEC,res}}_{t}
&= \sum_i \left(P^{\mathrm{con}}_{i, t} - P^{\mathrm{self}}_{i, t}\right).
\end{align}

The amount of electricity exchanged locally within the~\ac{LEC} is
\begin{equation}
P^{\mathrm{loc}}_{t}
= \min\!\left(P^{\mathrm{LEC,sur}}_{t},\, P^{\mathrm{LEC,res}}_{t}\right).
\end{equation}

Locally exchanged electricity is allocated to households with residual demand proportionally to their residual consumption:
\begin{equation}
P^{\mathrm{loc, buy}}_{i, t}
=
\begin{cases}
\dfrac{P^{\mathrm{con}}_{i, t} - P^{\mathrm{self}}_{i, t}}
      {P^{\mathrm{LEC,res}}_{t}} \, P^{\mathrm{loc}}_{t},
& P^{\mathrm{LEC,res}}_{t} > 0,\\[8pt]
0, & P^{\mathrm{LEC,res}}_{t} = 0.
\end{cases}
\end{equation}

According to the surplus PV generation allocation rules in Section~\ref{Internal pricing and allocation rules}, local electricity sales are attributed to~\ac{PV}-owning households proportionally to their surplus~\ac{PV} generation:
\begin{equation}
P^{\mathrm{loc,sell}}_{i, t}
=
\begin{cases}
\dfrac{P^{\mathrm{sur}}_{i, t}}
      {P^{\mathrm{LEC,sur}}_{t}} \, P^{\mathrm{loc}}_{t},
& P^{\mathrm{LEC,sur}}_{t} > 0,\\[8pt]
0, & P^{\mathrm{LEC,sur}}_{t} = 0,
\end{cases}
\end{equation}

Accordingly, household-level grid imports and exports under the~\ac{LEC} case are:
\begin{align}
P^{\mathrm{LEC,import}}_{i, t}
&= P^{\mathrm{con}}_{i, t}
- P^{\mathrm{self}}_{i, t}
- P^{\mathrm{loc, buy}}_{i, t},\\
P^{\mathrm{LEC,export}}_{i, t}
&= P^{\mathrm{sur}}_{i, t}
- P^{\mathrm{loc,sell}}_{i, t}.
\end{align}

\subsubsection*{Household energy cost under LEC}

The energy cost of household $i$ over the time horizon $T$ under the~\ac{LEC} case is given by:
\begin{equation}
\begin{split}
C^{\mathrm{LEC}}_{i} =
\sum_{t=1}^{T} \bigg(
& P^{\mathrm{loc,buy}}_{i, t} 
\left(\lambda^{\mathrm{loc}} + (1-\gamma)\lambda^{N} \right )
- P^{\mathrm{loc,sell}}_{i,t}\,\lambda^{\mathrm{loc}}  \\
& + P^{\mathrm{LEC,import}}_{i,t}\lambda^{\mathrm{import}}- P^{\mathrm{LEC,export}}_{i,t}\,\lambda^{\mathrm{export}} \bigg),\\
\end{split}
\end{equation}

where $\gamma = 40\%$ when local exchange occurs without a change in voltage level, or $20\%$ otherwise. $\lambda^{N}$ is the volumetric network usage fee.
\subsection{Evaluation metrics}

To quantitatively assess the impacts of the~\ac{LEC} framework, we evaluate system performance using a set of household-level and community-level metrics. All metrics are computed consistently for the reference case (no~\ac{LEC}) and the~\ac{LEC} case over the analysis horizon $T$.

\subsubsection*{Community self-consumption}
At the community level, the self-consumption rate is defined as the share of total~\ac{PV} generation that is consumed within the community, including both household self-consumption and local electricity exchange:
\begin{equation}
\mathrm{SCR}
=
\frac{\sum_{t=1}^{T} (P^{\mathrm{loc}}_{t} + \sum_i P^{\mathrm{self}}_{i, t})}
{\sum_{t=1}^{T} \sum_i P^{\mathrm{gen}}_{i,t}}.
\end{equation}
\subsubsection*{Local electricity exchange}
Local electricity exchange can be evaluated from two complementary perspectives: the utilization of local generation and the coverage of local demand. To capture both aspects, we define two local exchange rate indicators.

The generation-based local exchange rate quantifies the share of total local electricity generation that is exchanged within the community rather than exported to the grid:
\begin{equation}
\mathrm{LER_\mathrm{gen}}
=
\frac{\sum_{t=1}^{T} P^{\mathrm{loc}}_{t}}
{\sum_{t=1}^{T} \sum_i P^{\mathrm{gen}}_{i,t}}.
\end{equation}

The consumption-based local exchange rate measures the fraction of total electricity demand that is satisfied through local exchange:
\begin{equation}
\mathrm{LER_\mathrm{con}}
=
\frac{\sum_{t=1}^{T} P^{\mathrm{loc}}_{t}}
{\sum_{t=1}^{T} \sum_i P^{\mathrm{con}}_{i,t}}.
\end{equation}

\subsubsection*{Household electricity cost savings}

Household-level cost savings from participating in the~\ac{LEC} are computed as:
\begin{equation}
\Delta C_{i}
=
C^{\mathrm{ref}}_{i}
-
C^{\mathrm{LEC}}_{i}.
\end{equation}

Positive values of $\Delta C_{i}$ indicate net economic benefits from~\ac{LEC} participation.

\subsubsection*{Grid interaction}
Grid interaction is evaluated based on residual electricity exchange between the~\ac{LEC} and the distribution grid, considering both total energy volumes and peak power levels over the analysis horizon. Total imports and exports are defined as the aggregated electricity exchanged with the grid across all households and time steps, while peak interaction is captured by the maximum import and export levels. These metrics indicate, respectively, the overall reliance on the grid and potential impacts on capacity requirements and congestion risks.

\section{Results and discussions} \label{Results analysis}

\subsection{Input data}
Based on the modeling framework introduced in Section~\ref{modeling}, a~\ac{LEC} system is analyzed using a real-world demonstrator, namely the NEST platform~\cite{nest2026empa}. NEST is a modular research and innovation building located on the Empa campus in Dübendorf, Switzerland, serving as a representative vertical energy district. The case study comprises seven units, including residential apartments, office spaces, and a fitness facility. Among these, three units are equipped with~\ac{PV} systems, while the remaining units act as pure consumers. Detailed information on installed PV capacity, annual generation, and electricity demand is summarized in Table~\ref{tab:Installed PV capacity, annual generation, and demand in 2025}.
\begin{table}[ht]
\caption{Installed PV capacity, annual generation, and demand in 2025}
\label{tab:Installed PV capacity, annual generation, and demand in 2025}
\centering
\footnotesize
\setlength{\tabcolsep}{3.5pt}
\begin{tabularx}{\linewidth}{l *{7}{>{\centering\arraybackslash}X}}
\toprule
Unit ID & 1 & 2 & 3 & 4 & 5 & 6 & 7 \\
\midrule
PV cap. (kWp)   & 7.3  & 2.1  & 26.7  & 0 & 0 & 0 & 0 \\
PV gen. (kWh)  & 4418 & 1060 & 16760 & 0 & 0 & 0 & 0 \\
Demand (kWh)  & 4682 & 3291 & 15863 & 2269 & 2264 & 3511 & 12195 \\
\bottomrule
\end{tabularx}
\end{table}

Electricity prices are based on a representative retail tariff structure for low-voltage residential customers in the Zürich region in 2026. The average retail tariff for grid-imported electricity is assumed to be 0.241~CHF/kWh~\cite{ekz2026standardtariff}, while surplus~\ac{PV} generation exported to the public grid is remunerated at a feed-in tariff of 0.06~CHF/kWh~\cite{ekz2026feedintariff}. In addition, electricity imports are subject to a volumetric network usage fee of 0.0859~CHF/kWh and non-reducible levies amounting to 0.0344~CHF/kWh.

Within the LEC, internal electricity exchange prices are assumed to vary between 0.06 and 0.12~CHF/kWh, spanning the interval between the grid feed-in tariff and the average retail energy price. A network charge reduction factor of $\gamma = 0.4$ is applied to locally exchanged electricity, reflecting local exchanges that do not involve a change in voltage level. 

\subsection{Community-level impacts of LECs}

At the community level, the~\ac{LEC} configuration leads to a substantial improvement in the utilization of local renewable generation. The self-consumption rate increases from 39.4\% in the reference case to 62.3\% under the~\ac{LEC} case, indicating a more effective alignment between local generation and local demand. From the generation perspective, 22.2\% of the locally produced electricity is consumed via local exchange, while from the demand perspective, 11.6\% of total electricity consumption is supplied by locally shared~\ac{PV} generation.

Beyond energy balance metrics, the~\ac{LEC} also induces meaningful changes in grid interaction. Total electricity imports and exports are reduced by 14.4\% and 37.8\%, respectively, compared to the reference case. Importantly, peak import demand remains unchanged at 10.8~kW in both cases, suggesting that peak demand events predominantly occur during periods with limited or no PV generation. In contrast, peak export power decreases by approximately 13\%, from 15.5~kW to 13.5~kW.

From a system perspective, these results indicate that~\ac{LEC} primarily acts as a renewable integration and export-smoothing mechanism, rather than a peak demand reduction measure in the absence of storage. While the impact on import peaks is limited, the reduction in export peaks can alleviate local grid congestion risks and reduce reverse power flow stresses as distributed~\ac{PV} penetration continues to rise.

\subsection{Household-level distributional effects}
\begin{figure}[!htbp]
    \centering  \includegraphics[width=0.9\linewidth]{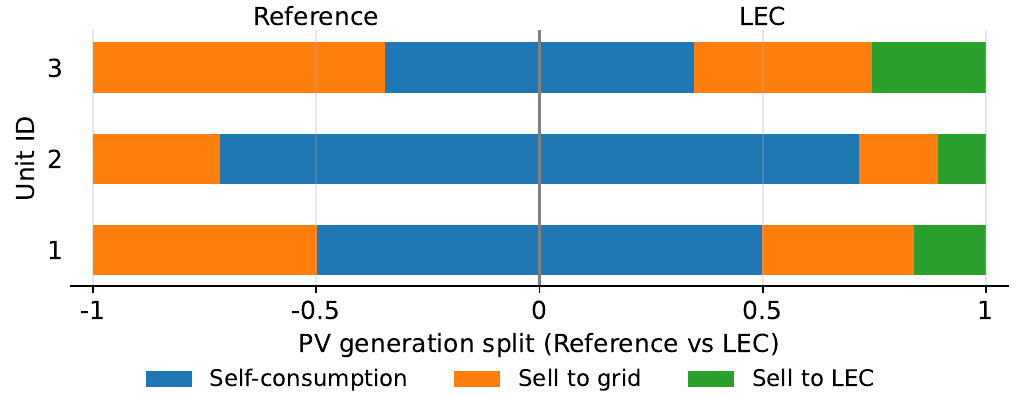}
    \caption{PV generation split under the reference (left) and LEC (right) cases. Colors indicate where the PV generation is allocated.}
    \label{fig:PV split}
\end{figure}

Figure~\ref{fig:PV split} illustrates how~\ac{PV} generation is allocated among self-consumption, grid export, and local exchange for~\ac{PV}-equipped households under both the reference and~\ac{LEC} cases. A key observation is that approximately 38\% of the~\ac{PV} electricity that would otherwise be exported to the grid in the reference case is absorbed locally within the~\ac{LEC}. This demonstrates the effectiveness of local electricity sharing in increasing the local utilization of distributed~\ac{PV} generation without requiring additional physical infrastructure.

\begin{figure}[!htbp]
    \centering
\includegraphics[width=0.9\linewidth]{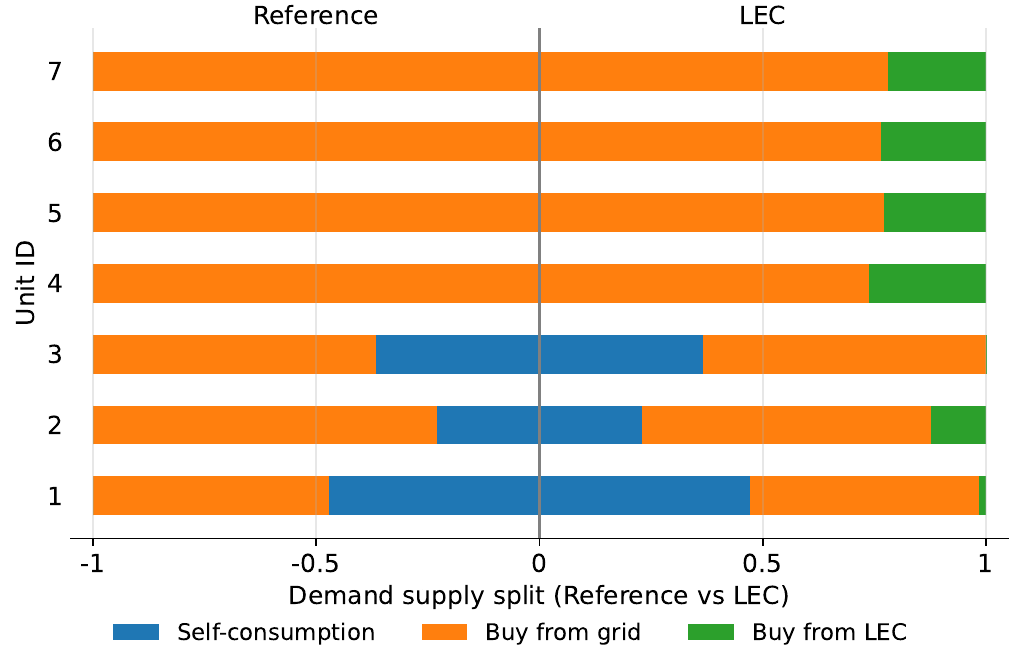}
    \caption{Shares of self-consumption, grid imports, and local exchange in meeting household electricity demand under reference (left) and LEC (right) cases. Colors indicate where the supply is from.}
    \label{fig:demand split}
\end{figure}

Fig.~\ref{fig:demand split} further decomposes electricity supply sources from the demand side. In the reference case, non-\ac{PV} units rely entirely on grid imports, while units with~\ac{PV} meet their demand through a combination of self-consumption and grid electricity due to the temporal mismatch between~\ac{PV} generation and demand. Under the~\ac{LEC} configuration, non-PV units benefit from access to locally shared~\ac{PV} electricity, while~\ac{PV} units continue to rely primarily on self-consumption and grid imports.

An interesting asymmetry emerges among PV units. Units with relatively small PV systems (e.g., Unit~2, whose annual PV generation covers only about one-third of its annual electricity demand, as shown in Table~\ref{tab:Installed PV capacity, annual generation, and demand in 2025}) still rely substantially on local exchange, and behave more like consumers than surplus-generating prosumers. In contrast, units whose annual PV generation is closer to their annual demand exhibit limited reliance on local exchange. This highlights that PV sizing relative to demand plays a critical role in determining participation patterns and benefit distribution within a~\ac{LEC}, and should therefore be assessed not only at the individual level but also in the context of the community.

\subsection{Economic distribution and pricing fairness}
\begin{figure}[!htbp]
    \centering
    \includegraphics[width=0.9\linewidth]{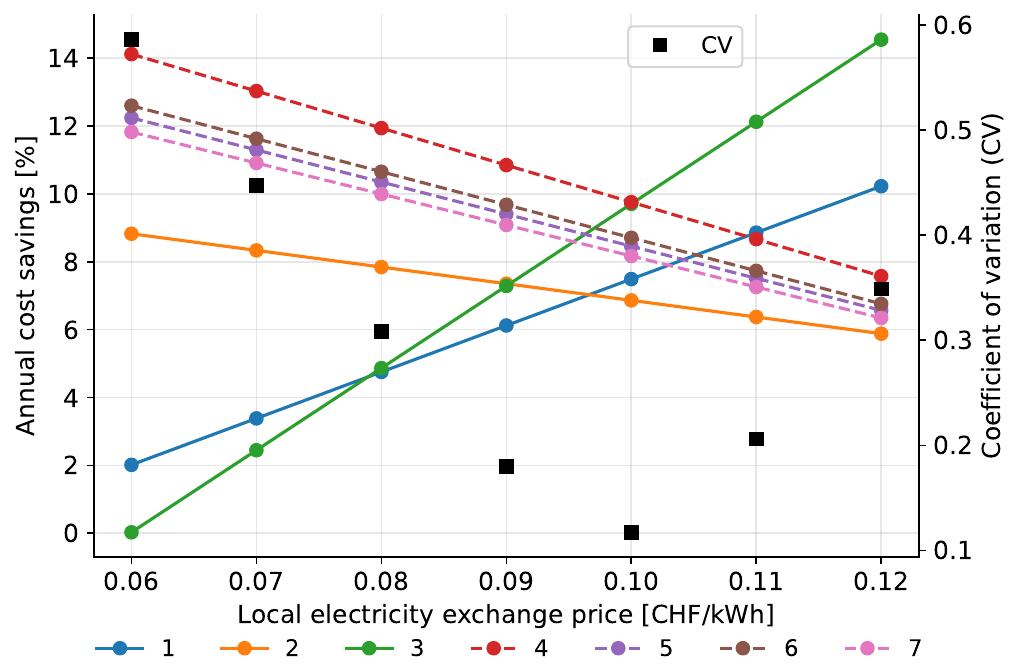}
    \caption{Sensitivity analysis: annual cost savings VS local electricity exchange price, dash lines present units with PV, dotted line present units without PV.}
    \label{fig:sensitivity analysis}
\end{figure}

To assess the sensitivity of household-level economic outcomes to the internal pricing of local electricity exchange, an analysis is conducted with local exchange prices ranging from 0.06 to 0.12~CHF/kWh. As shown in Fig.~\ref{fig:sensitivity analysis}, households with PV systems experience increasing cost savings as the local exchange price rises, reflecting higher revenues from selling locally rather than exporting to the grid. Conversely, non-PV households benefit more from lower local exchange prices, as they gain access to electricity below standard retail tariffs.

Households with limited PV capacity (e.g., Unit~2) exhibit intermediate behavior, reinforcing the earlier observation that PV sizing strongly shapes economic roles within the community. These results reveal a distributional trade-off embedded in~\ac{LEC} pricing: prices that favor prosumers may disadvantage consumers, and vice versa.

To identify a pricing level that balances economic benefits across participants, the~\ac{CV} of household cost savings (defined as the standard deviation relative to the mean of saving percentages) is evaluated (Fig.~\ref{fig:sensitivity analysis}, black squares). This metric captures the equity of benefit distribution: lower values indicate more uniform savings, while higher values reflect greater disparities among participants. The minimum~\ac{CV} is observed at a local exchange price of approximately 0.10~CHF/kWh, suggesting that this price yields the most equitable distribution of benefits for the specific~\ac{LEC} configuration studied. This analysis provides a quantitative basis for supporting internal price negotiations within~\acp{LEC}, complementing regulatory flexibility with data-driven decision support.

\subsection{Discussions}
The results indicate that~\ac{LEC} participation provides economic benefits to both consumers and prosumers: consumers gain access to locally generated electricity at prices below retail tariffs, while prosumers obtain higher revenues than grid feed-in remuneration. At the community level,~\acp{LEC} increase local renewable utilization and reduce electricity exports and export peaks, supporting renewable integration, although peak import demand remains largely unchanged without additional flexibility.

This analysis is based on a single real-world case and does not include storage or operational costs. Nonetheless, it offers a practical, data-driven reference for how~\acp{LEC} may perform in practice and provides relevant insights for households,~\ac{LEC} operators, and~\acp{DSO} considering the deployment and scaling of energy communities. It also offers a fair internal pricing mechanism for both prosumers and consumers within the~\ac{LEC}. 
\par

\section{Conclusions} \label{Conclusions}
With the Swiss~\ac{LEC} framework entering into force on 1 January 2026, this study provides a forward-looking assessment of how~\ac{LEC} participation may affect local energy use and household electricity costs in practice. Using empirical data from the NEST demonstrator and vertical energy district as a case study, the analysis shows that local electricity sharing can significantly increase the use of locally generated renewable electricity, reduce grid exports, and generate economic benefits for both consumers and prosumers. Beyond efficiency gains, the analysis highlights that internal electricity sharing pricing plays a critical role in shaping the distribution of economic benefits between households with and without~\ac{PV}, making price design a key practical consideration for equitable and effective ~\ac{LEC} implementation. Overall, the findings provide actionable insights for households,~\ac{LEC} coordinators, and~\ac{DSO} in preparing for the deployment of~\ac{LEC}s in Switzerland. Future work will extend the framework to incorporate energy storage, additional energy conversion technologies, and alternative tariff structures, in order to further assess their potential to enhance~\ac{LEC} performance and flexibility.

\balance

\tiny
\bibliographystyle{IEEEtran}
\bibliography{ref}

\end{document}